# Search for New Charged Gauge Boson Wprime via Phenomenology of the Left-Right Symmetric Model at Hadron Colliders

Nady Bakhet[1,2,5], Tarek Hussein[1,6], Maxim Yu. Khlopov[3,4,7]

[1]Physics Department-Faculty of Science-Cairo University-Egypt
[2]Egyptian Network of High Energy Physics –ASRT-Egypt
[3]APC Laboratory, IN2P3/CNRS, Paris, France
[4]National Research Nuclear University "Moscow Engineering Physics Institute"
115409 Moscow, Russia
[5]nady.bakhet@cern.ch, [6]tarek@sci.cu.edu.eg, [7]khlopov@apc.univ-paris7.fr

## Abstract

In this article, a search for signatures of the hypothetical new heavy charged gauge boson Wprime in proton-proton collisions at the Large Hadron Collider (LHC) with center-of-mass energy range from 6 to 14 TeV and in proton-antiproton collisions at the Fermilab Tevatron Collider with center-of-mass energy of 1.96 TeV. The production of W' and its decay were simulated via the decay channel to top and anti-bottom by using Monte Carlo events generator programs. The events were simulated according to the extension of Standard Model (SM) and the Left-Right symmetry Model (LRSM) based on the gauge symmetry group $SU(3)_C \times SU(2)_L \times SU(2)_R \times U(1)_{B-L}$. The LRSM model exhibits signatures of new physics Beyond the Standard Model (BSM) at the hadron colliders. The most interesting arising from the LRSM is the production and decay of new boson Wprime. We present analysis of detecting the signal of Wprime bosons using the process $pp(\bar{p}) \to W' \to t\bar{b} \to \ell \nu b \bar{b}$ in final state with (electrons or muons) and missing transverse energy. The production of Wprime boson was found at the LHC at center of mass energy 10 TeV with mass 1TeV and the production cross section of 13.31pb.

# I. INTRODUCTION

The main purpose of the hadron colliders is to search for signals of new physics beyond the Standard Model (SM). The SM can not present an explanation for many fundamental problems such as the hierarchy problem (resulting from the large difference between the weak force and the gravitational force) and dark matter. It is, therefore, widely believed that new physics beyond the SM will be discovered in the coming years. At current energies, the world is left-handed where the Standard Model contains an SU(2)L group.

Among the possible attractive platform for new physics is the extension of the SM, the Left-Right Symmetric Model (LRSM) based on the gauge symmetry group SU(3)C × SU(2)L × SU(2)R × U(1)B-L  [1,2]. This gauge group can be understood as a second step after the SM to unify the fundamental interactions and solve several problems of the SM.  Left-right symmetry at some larger scale implies the need for an SU(2)_R group. Thus the particle content is expanded by right-handed Z' and W' and right-handed neutrinos. In the SM both left-handed neutrinos and charged leptons of the same flavour are elements of the same object with respect to the SU(2) gauge group and also the left handed quarks. However, right-handed fermions fields are not treated in the same way they are singlet fields with respect to the SU(2) gauge group. In the left-right symmetric model both left-handed and right-handed fields they form doublets under $SU(2)_L$ and $SU(2)_R$ gauge groups. LRSM try to find  a solutions  for  the problems of the SM, Parity violation in the weak interactions, and non-zero neutrino masses implied by the experimental evidence of neutrino oscillation. The left-right symmetry which underlies LRSM restores parity symmetry at energies higher than the electroweak (EW) scale, resulting in the new heavy gauge bosons, W' and Z'. LRSM exhibits the neutrinos are heavy where their nature i.e., whether they are of  Majorana or Dirac type depends on the details of the LRSM. The Higgs sector in the previous concept of the LRSM contains a Higgs bidoublet and two Higgs doublets so the neutrinos are of Dirac type and no natural explanation for their small masses. Now, LRSM incorporates a Higgs bidoublet and two Higgs triplets which lead to Majorana type neutrinos. The LRSM provides an explanation for the smallness of neutrino (Majorana) masses, relating their mass scale to the large left-right symmetry breaking scale through the see-saw mechanism.

The LRSM [3,4] is based on the gauge group $SU(3)_C \times SU(2)_L \times SU(2)_R \times U(1)_{B-L}$ under which the fermions are doublets.

$$L_{iL} = \begin{pmatrix} \nu_i' \\ \ell_i' \end{pmatrix}_L : (2,1,-1), \quad L_{iR} = \begin{pmatrix} \nu_i' \\ \ell_i' \end{pmatrix}_R : (1,2,-1)$$

$$Q_{iL} = \begin{pmatrix} u_i' \\ d_i' \end{pmatrix}_L : (2,1,\frac{1}{3}), \quad Q_{iR} = \begin{pmatrix} u_i' \\ d_i' \end{pmatrix}_R : (1,2,\frac{1}{3})$$

i=1,2,3 runs over number of generations. The numbers ($d_L$, $d_R$, Y) in parenthesis characterize the $SU(2)_L$, $SU(2)_R$ and $U(1)_{B-L}$ representation. $d_{L,R}$ denote dimensions of the $SU(2)_L$ and $SU(2)_R$ representation, Y = B − L. The quantum numbers for $U(1)_{B-L}$ gauge group are connected with charges of the particles

We implemented the Left-Right Symmetric Model (LRSM) by FeynRules package to translate the Lagrangian of LRSM model to all the interaction vertices of the model and to face the output to any Feynman diagram calculators MadGraph/Madevent or Calchep. We put Kinematics cuts on Monte Carlo event generators PYTHIA8 [5-8] which also used to showering and hadronization then we use ROOT data analysis and Physical Analysis Workstation (PAW) for analyzing and studying the new physics signals of W' gauge boson from Monte Carlo data generated at the LHC and Tevatron and finally. We produced different dynamics of W' gauge boson.

Our work is organized as follows. In the second section we present the results of numerical calculations produced from our simulation for Left-Right Symmetric Model for predication and decay of W' boson, the first subsection of these results, we calculated the production cross section, branching ratios, width, reconstruction of W' boson mass from Invariant mass of top-bottom quarks, also we calculated the transverse momentum and pseudorapidity for top and bottom quarks. The second subsection of the results describes the detection of W' signal at the hadron colliders via electron or muon decay channel in final state where we calculated the production cross section, transverse momentum and pseudorapidity for both electron and muon at the LHC and Tevatron. In the third section the summary and conclusion are given.

## II. The Results

In this section, we present the results of our analysis for the hypothetical W' boson a new heavy charged gauge boson that decays to a top-bottom quarks at the hadron colliders, the Tevatron and LHC by using Mont Carlo events generators Pythia8, Calchep [9] and MadGraph5 [10]. Also we will focus on the production of heavy charged gauge boson via the decay channel of charged leptons (electron or muon) plus missing transverse energy in final state where W′ boson decay has a good separation between the W′ boson signal and the SM backgrounds. Search for right-handed W' bosons that decay to a top-bottom quarks W'→ tb (or charge conjugate) make no assumptions regarding the mass of the right-handed neutrino. The interactions of charged weak currents are realized in SM via exchange of charged massive gauge boson fields W+ and W−. Although any additional charged massive bosons are not found yet experimentally their existence is predicted by various extensions of SM. The wide-common name for this vector boson field is W′. One of these models is Left-Right Symmetric Model (LRSM). In some models the W' boson may couple more strongly to fermions of the third generation than to fermions of the first and second generations [11]. Thus the W'→ tb decay is an important channel in the search for W' bosons. Experimental searches for W→tb decays have been performed at the Tevatron [12] and at the LHC [24, 25]. The CMS search at Sqr(s )= 7 TeV [13] set the best present mass limit in this channel of 1.85 TeV for W' bosons with purely right-handed couplings. Frequently W′ bosons are discussed in connection with so called Left-Right symmetric models [14, 15].

The searches for W' bosons with purely right-handed couplings have been performed by the CMS and ATLAS Collaborations assuming the mass of the right-handed neutrino of is less than the mass of the W' boson [16].

We analyzed the Mont Carlo data events at the LHC at different center of mass energies from 6 TeV to 14 TeV also at the Fermilab Tevatron Collider at center of mass energy of 1.96 TeV via final state invariant mass distribution of top and bottom quarks. The top and bottom quarks are used for search about new physics related to electroweak symmetry breaking at the Hadron colliders .

Previous direct searches for W' bosons in the quark decay channel have excluded the mass range below 261 GeV [17] and between 300 GeV and 420 GeV [18]. The W' boson lower mass limit in this decay channel is 786 GeV [19]. The SM W boson from the top quark decay then decays leptonically or hadronically. W' boson contributes to the top quark decay, but that contribution is negligible, again because of the large W' boson mass.

In our analysis for LRSM the W' boson has right-handed interactions, hence there is no interference with the SM W boson where mixing angle very small. Also the W' boson is allowed to decay both to leptons and quarks,

The top quark was discovered in 1995 by the CDF and DØ collaborations [20], but the production of single top quark has not yet been observed. Both collaborations have searched for single top quark production [21]. Production of top-bottom final state via a $W^{'}$ right-handed boson does not interfere with top-bottom quarks production via a W boson and therefore the $W'$ sample only includes W' production. For all simulated samples, Pythia8 is used for parton showering, hadronisation and simulation of the underlying event. The Pythia8 and Madgraph5 backgrounds use the CTEQ6L1 PDFs,

The distinguishing feature of a W' signal is a narrow resonance structure in the top-bottom invariant mass spectrum.

# 1. Production of W' boson at the hadron colliders

## 1.1 Production Cross Section of W'

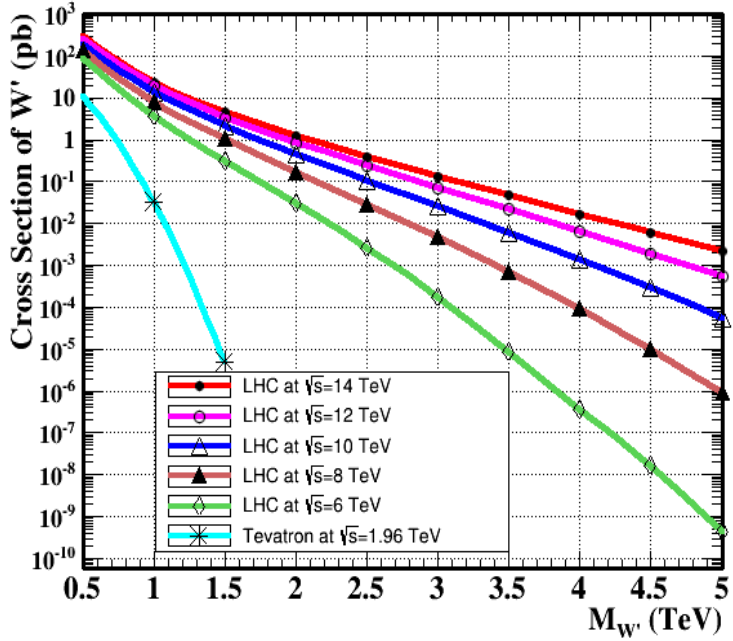

**Fig. 1.** The Production Cross Section of W' boson at the LHC for 6 to 14 TeV and at the Tevatron for different values of W' mass using MadGraph5/MadEvent

The theoretical W′ boson production cross section is more than 15 pb for masses between 200 GeV and 400 GeV for all three models considered here [22].

The current limits on the single top quark production cross section in the s-channel are 6.4 pb [23] and 13.6 pb [24] and don't depend much on whether the W boson coupling is left-handed or right-handed. Thus W′ boson production with decay to a top and a bottom quark is excluded in this mass region. In this analysis we therefore explore the region of even higher masses. From table 1 the production cross section of W' boson in the s-channel when its mass is 1TeV is 0.03297 pb at the Tevatron and at the LHC is 8.04 pb and 13.31pb at center of mass energy 8TeV and 10TeV respectively.

| Center of mass energy | LHC (TeV) | | | | | Tevatron (TeV) |
|---|---|---|---|---|---|---|
| | 14 | 12 | 10 | 8 | 6 | 1.96 |
| Cross Section (pb) for W' mass 1 TeV | 25.53 | 19.16 | 13.31 | 8.09 | 3.665 | 0.03297 |
| Cross Section (pb) for W' mass 1.5 TeV | 4.928 | 3.438 | 2.119 | 1.048 | 0.3148 | 4.761E-06 |

**Table 1.** Production Cross Section of W' at the LHC and Tevatron.

From table 1 we note that the production cross section of W' boson at the LHC for center of mass energy 10 TeV is 13.31 pb when the mass of W' is 1TeV this value is the nearest one for the theoretical value 15 pb [22] also this is near from the current limit on the single top quark production cross section in the s-channel which is 13.6 pb [24]. Also from table 1 the production cross of W' at 8TeV when the mass of W' is 1TeV is 8.009 pb and also near from the current limit on the single top quark production cross section in the s-channel which is 6.4 pb [23]. So we expect the production of top quark may come from the new charged heavy gauge boson W'.

## 1.2 Width of W' boson

The single top quark final state is sensitive to the presence of new charged heavy boson owing to the decay chain W′ → tb, where the top quark decays to a b quark and a SM W boson. This decay is kinematically allowed as long as the W′ mass is larger than the sum of top and bottom quarks masses, i.e. as long as it is above about 200 GeV.

The total width of the W′ in dependence on W′ mass for the top quark mass $M_t = 172$ GeV assuming that decays to both quarks and leptons are allowed

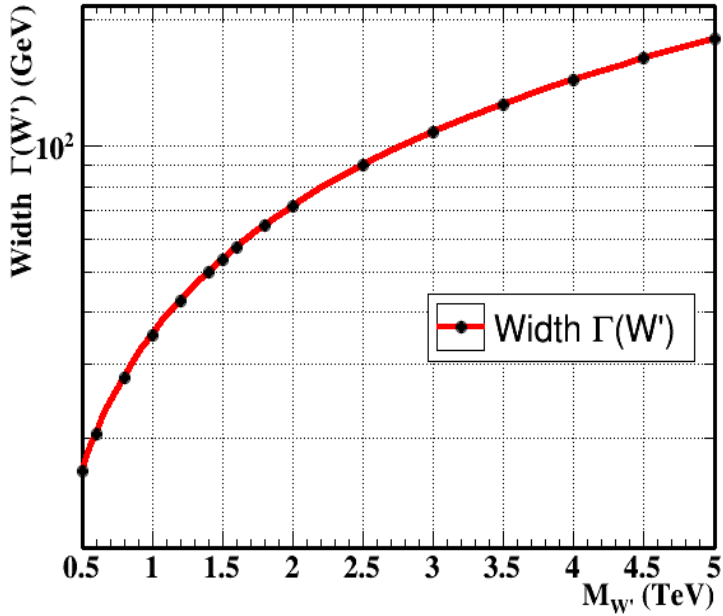

**Fig. 2.** The total width of the W' for different masses of W' and for top quark mass 172 GeV and heavy neutrino mass 100 GeV using CalcHep

## 1.3 Branching Ratios of W' boson

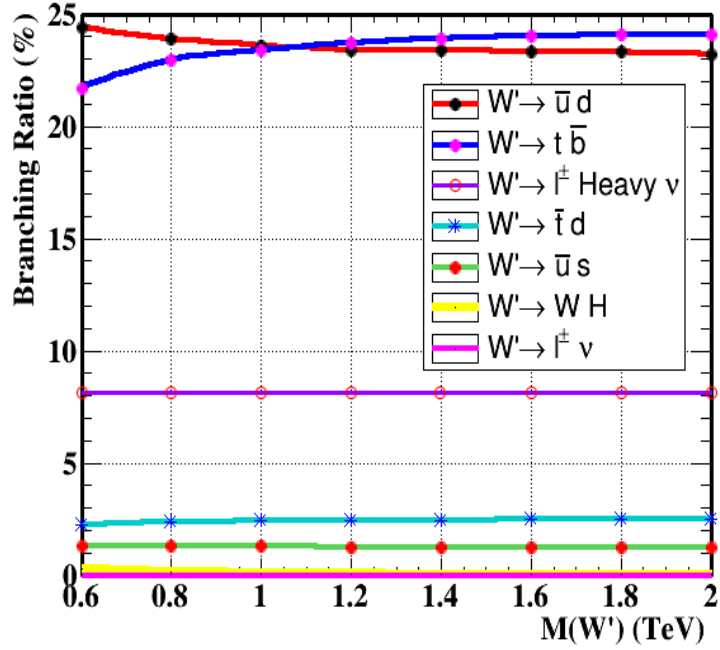

**Fig. 3:** Branching ratios of all decay channels of the W' for different masses of W' and top quark mass 172 GeV and heavy neutrino mass 100 GeV using CalcHep

Figure 3 shows all decay channels of W' boson and the decay channel to top-bottom quarks $W'_R \to tb$ has the high branching ratio than the other channels so this channel is used to detect the W' at the hadron colliders. The $W'$ boson can only decay leptonically if there is a right-handed neutrino $v_R$ of sufficiently small mass, M($v_R$), so that M($v_R$)+M($\ell$) < M(W'). If the mass of the right-handed neutrino is too large, $W'_R$ bosons can only decay to $q\bar{q}'$ final states, leading to different branching fractions for the $W'_R \to tb$.

## 1.4 Production Cross section of Top-Bottom from W' boson

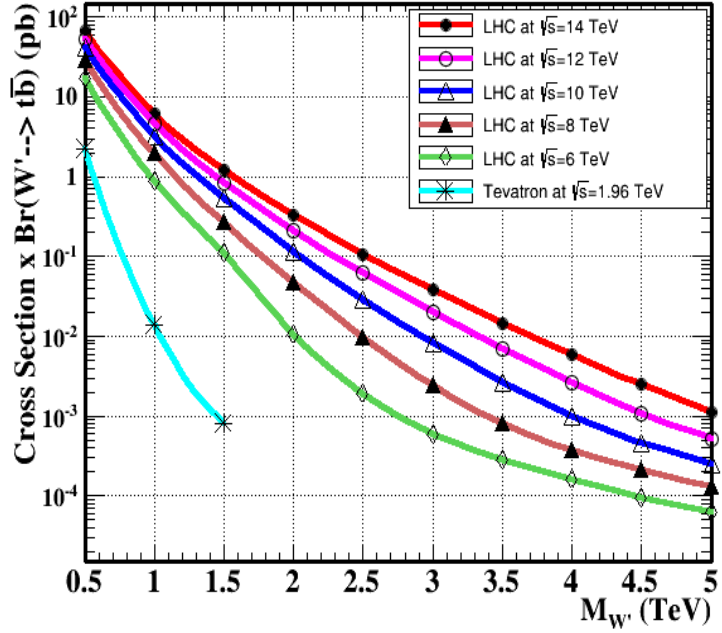

**Fig. 4:** The Production Cross Section of top-bottom via W' boson decay at the LHC for energies 6 to 14 TeV and at the Tevatron for different values of W' mass using MadGraph5/MadEvent

From figure 4 the production cross of top-bottom quarks to total production cross of W' is high because this channel has the branching fraction. For W' boson interactions, the product of production cross section and branching fraction depends on whether the decay to leptons is allowed or not. The branching fraction for the decay $W' \to t\bar{b}$ is about 25% if the W' boson decay to quarks and leptons is allowed.

## 1.5 Reconstruction of W' mass from Top-Bottom quarks

Figure 5 shows the reconstruction W' boson mass from top-bottom invariant masses for different values of W' boson masses from .5 to 3TeV also the figure shows the histogram of invariant mass of top-bottom of the Standard Model where this histogram has the lowest value between all histograms and we note some histogram has a peak in the invariant mass distribution of the tb final state $pp \rightarrow W' \rightarrow t\bar{b}$. We will use the notation tb includes both final States $W'^+ \rightarrow t\bar{b}$ and $W'^- \rightarrow \bar{t}b$. Figure 6 shows the differential cross section for the s-channel single-top quark production [25] as a function of the invariant mass of top-bottom quarks for all curves start from the reaction threshold at about M($t\bar{b}$) ≈ 200 GeV. The numerical calculation and Monte Carlo simulations have been performed for the Tevatron and LHC energies √s = 1.96 TeV and from 6 to14 TeV and the top quark mass was chosen 172 GeV and W' boson masses was chosen from 0.5 up to 3TeV for the LHC and from .5 to 1TeV for the Tevatron and the partonic distribution functions CTEQ6l1 have been used. The couplings of W′ to SM-fields have been implemented into CACHEP which was used to compute the W′ width production cross sections, kinematical distributions (transverse momentum and pseudorapidity) and to generate events for different W′ boson masses using MadGraph5

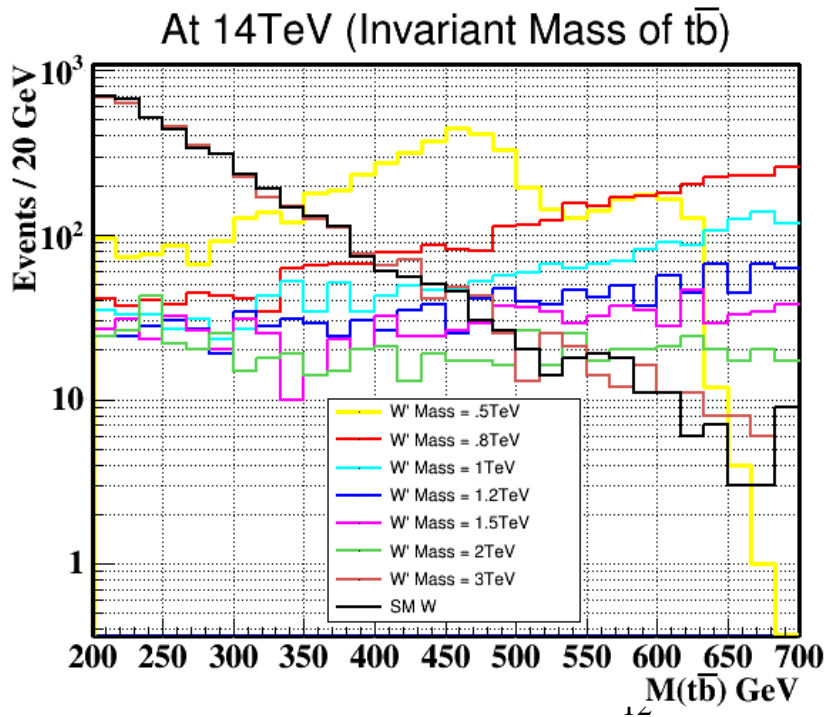

**Fig. 5:** The reconstruction W' mass boson from top-bottom invariant masses at the LHC at energy 14 TeV for different values of W' masses from .5 to 3 TeV also the invariant mass of the SM background top-bottom quarks.

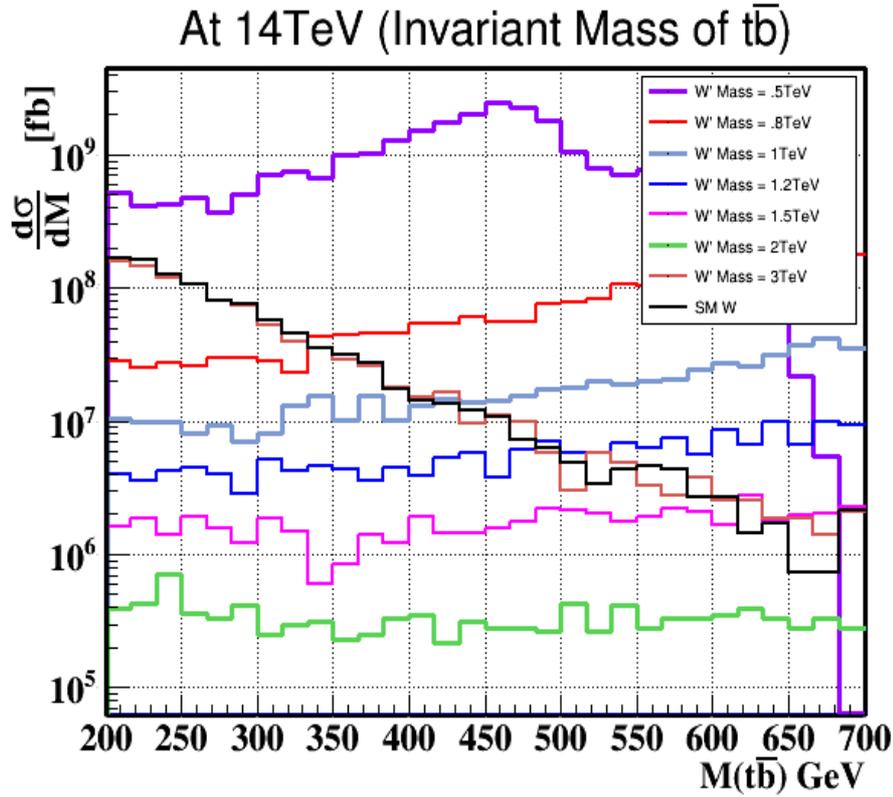

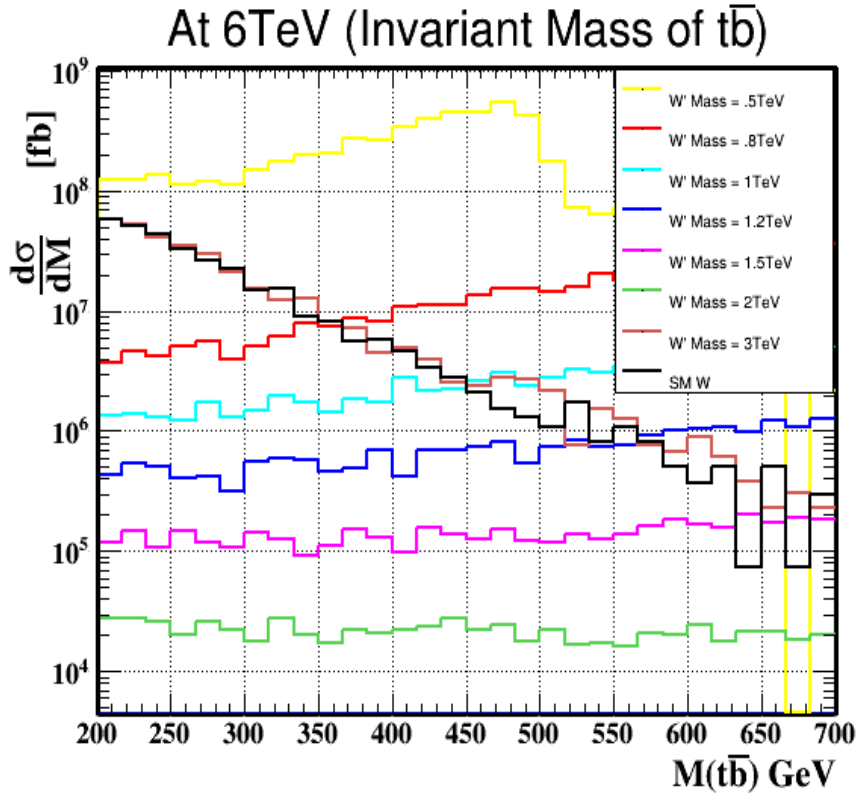

**Fig. 6:** The differential cross section for W' boson as a function of the invariant mass distribution of top-bottom for different mass values of W' boson 5 to 3 TeV at the LHC for 14 TeV (top) and 6 TeV (bottom) as well as SM background processes

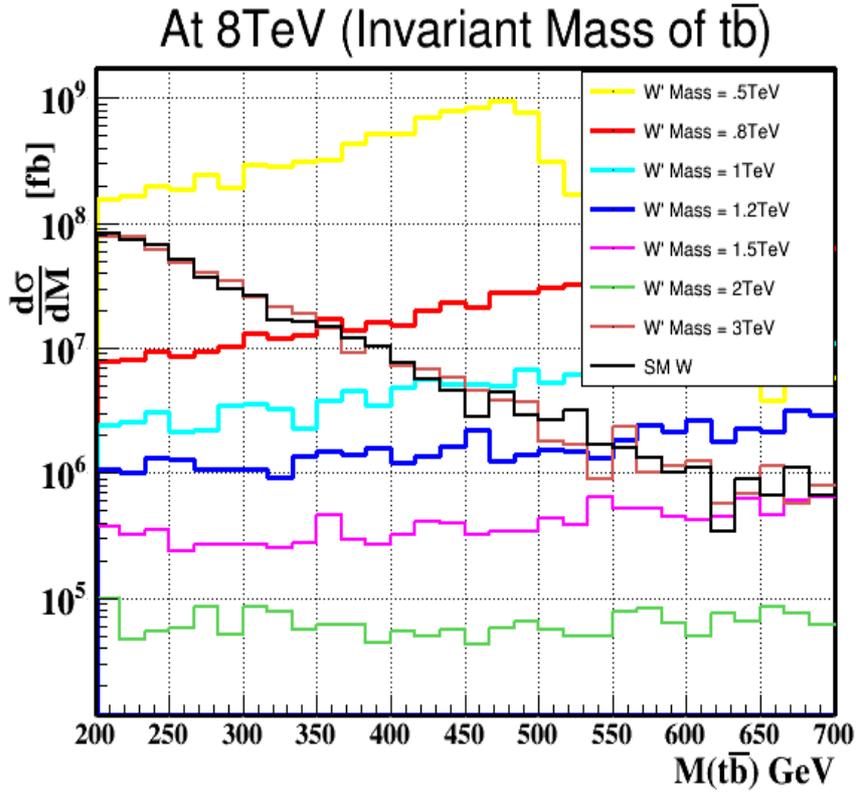

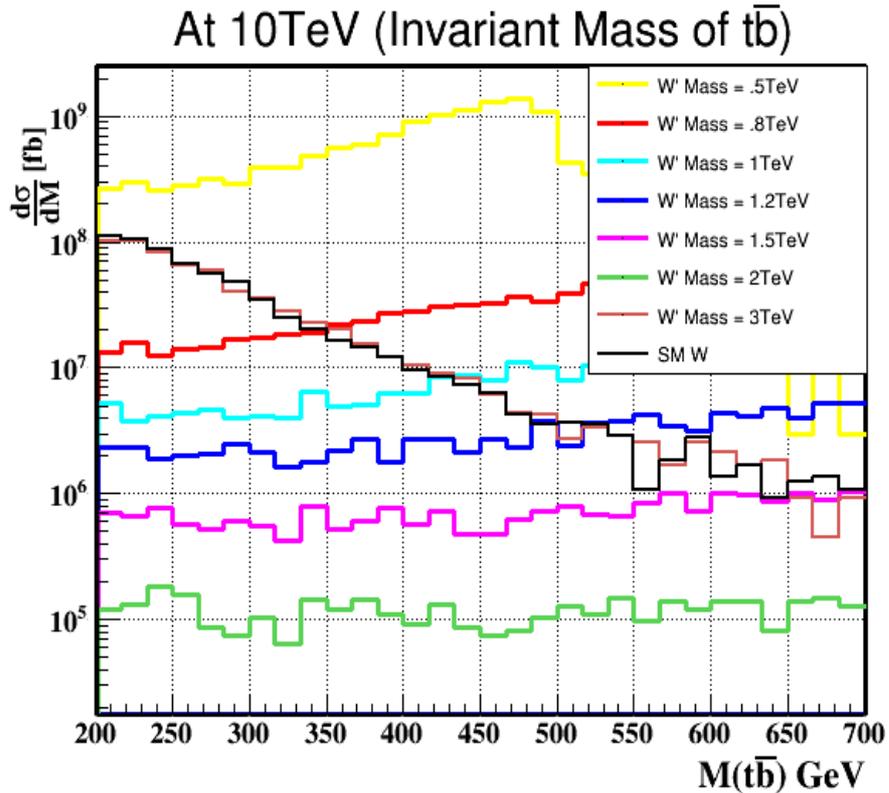

**Fig. 7:** The differential cross section for W' boson as a function of the invariant mass distribution of top-bottom for different mass values of W' boson 5 to 3 TeV at the LHC for 8 TeV (top) and 10 TeV (bottom) as well as SM background processes.

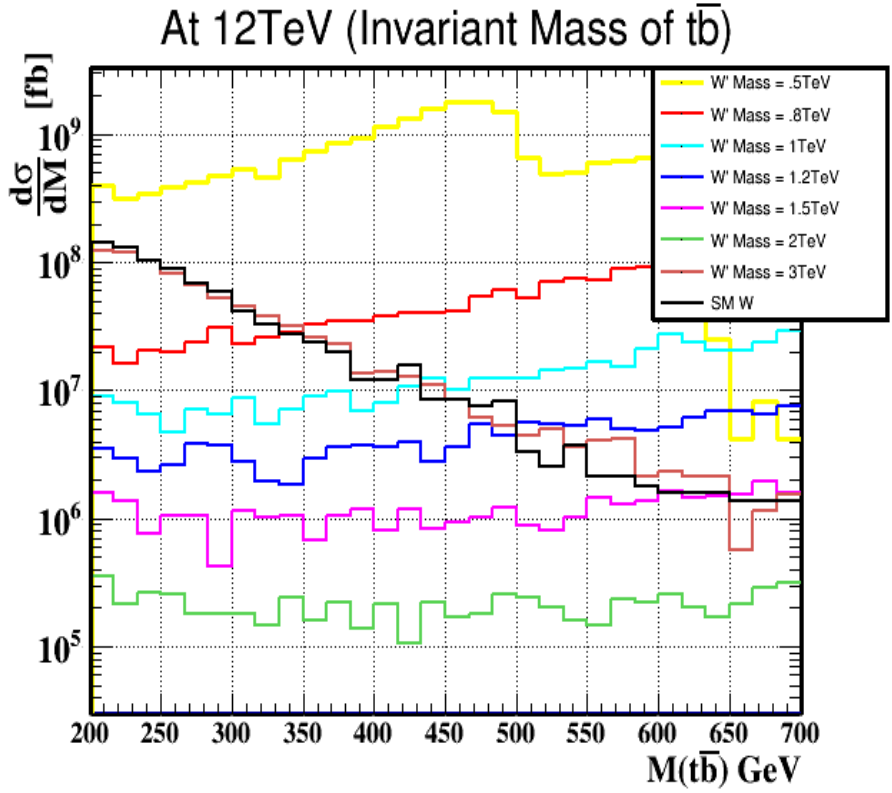

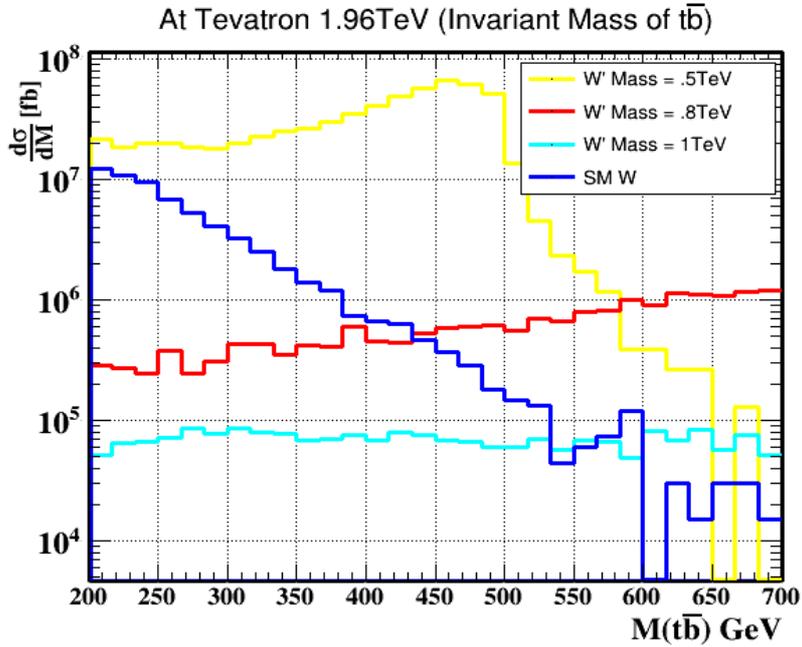

**Fig. 8:** The differential cross section for W' boson as a function of the invariant mass distribution of top-bottom for different mass values of W' boson 5 to 3 TeV at the LHC for 12 TeV and at Tevatron for W' boson mass .5 to 1TeV as well as SM background processes.

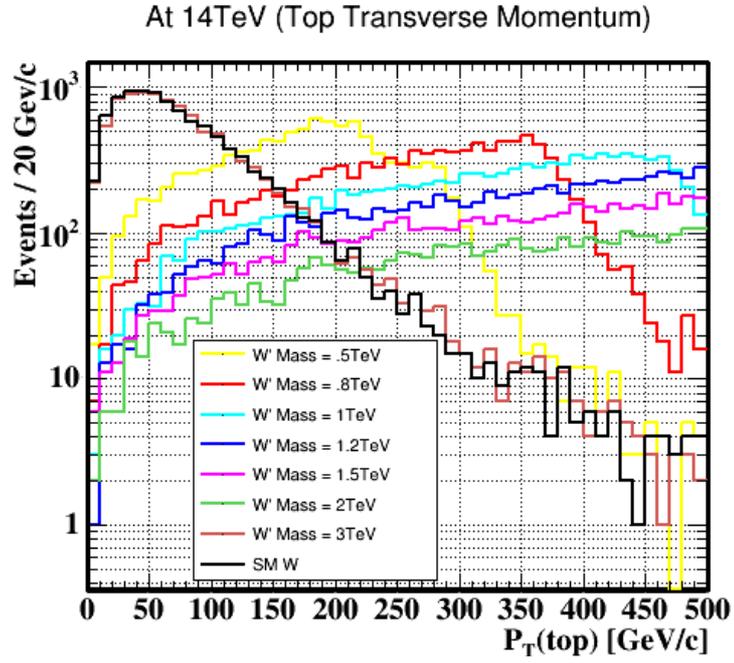

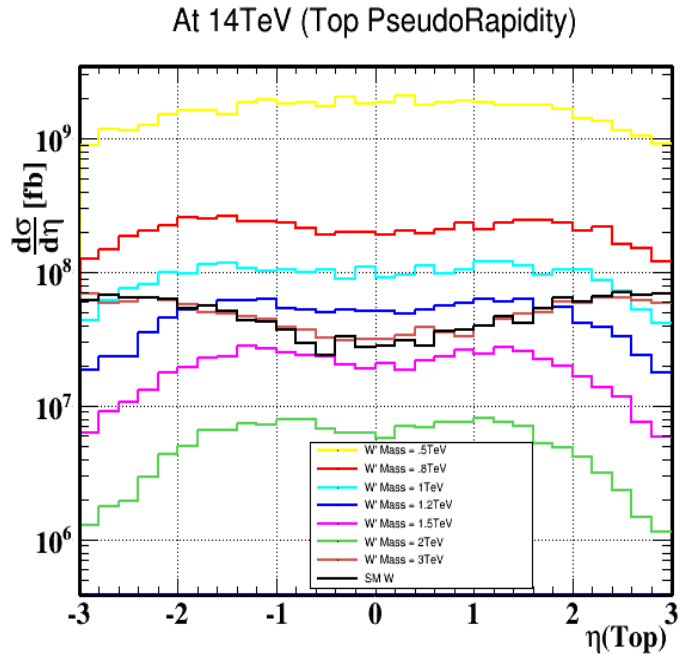

**Fig. 9:** The kinematics distributions of top quark production for different values of W' boson mass at the LHC at 14TeV, transverse momentum distribution and pseudorapidity distributions .

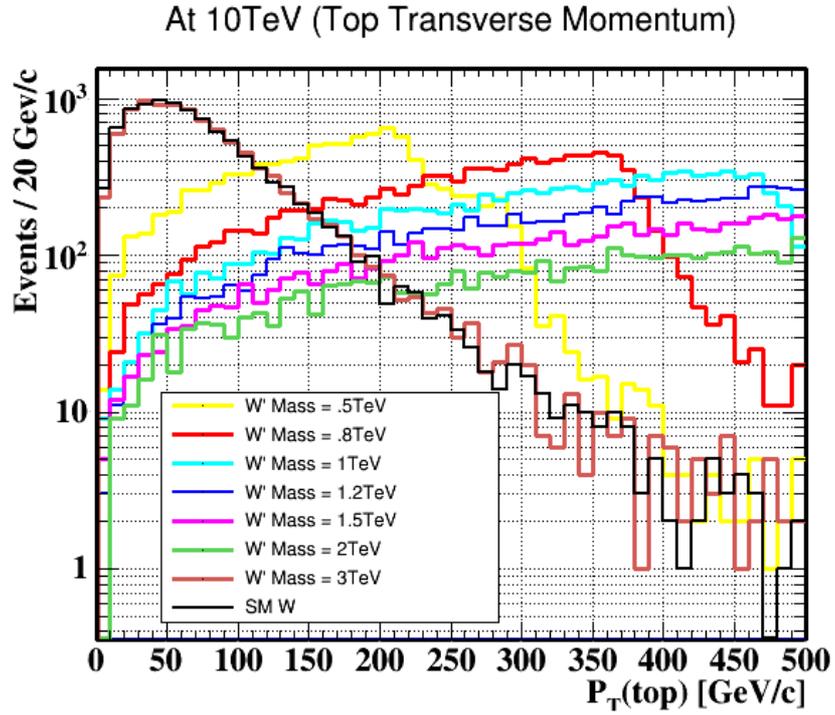

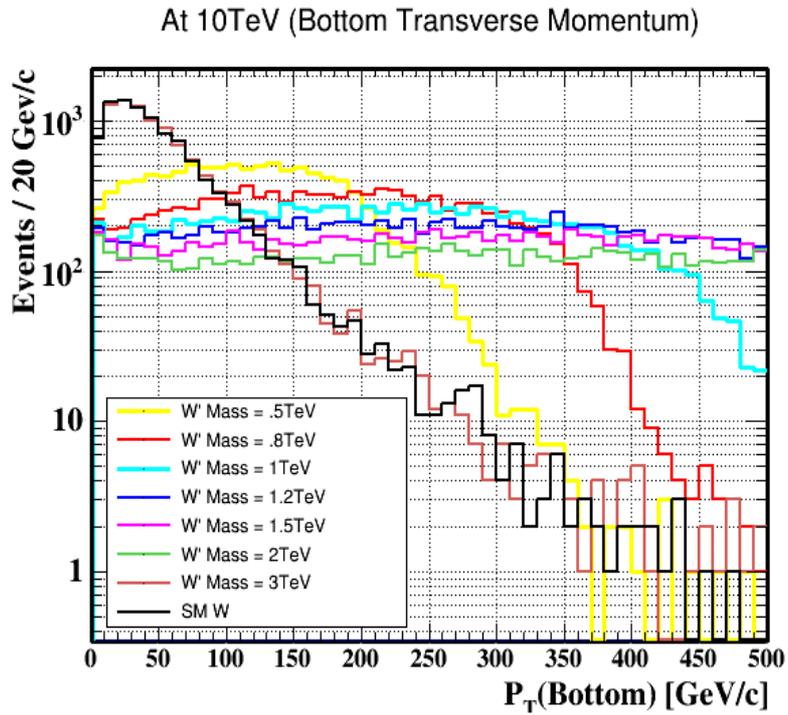

**Fig. 10:** The transverse momentum of top and bottom quarks for different values of W' boson masses at the LHC at 10 TeV

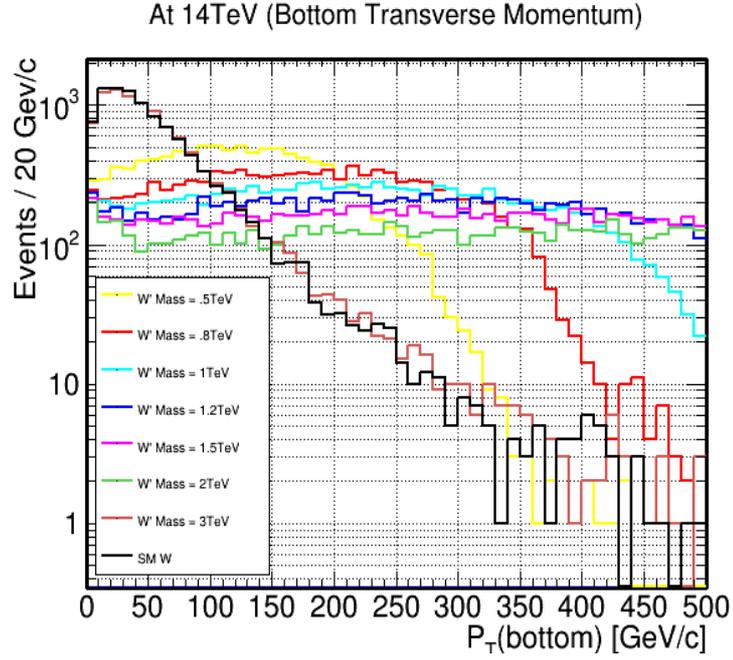

**Fig. 11:** The kinematics distributions of bottom quark production for different values of W' boson masse at the LHC at 14TeV, transverse momentum distribution and pseudorapidity distributions

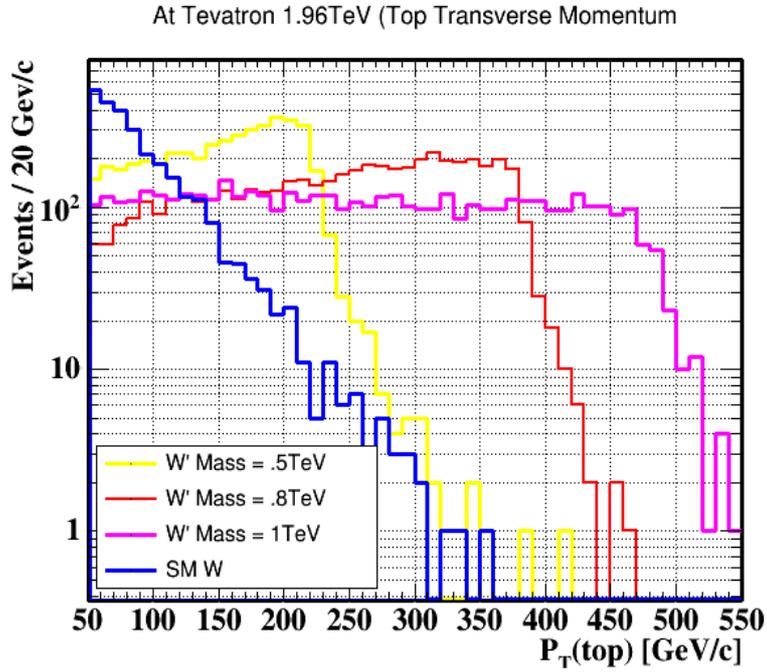

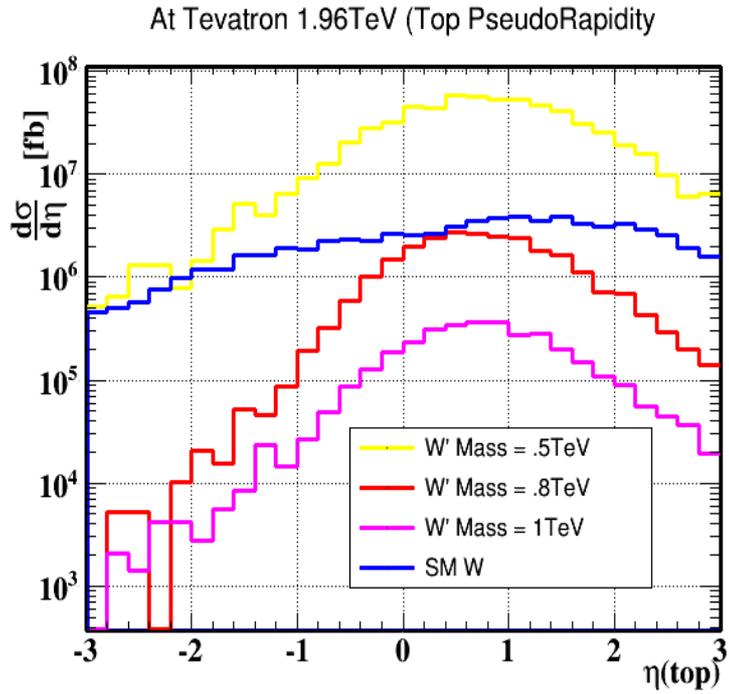

**Fig. 12:** The kinematics distributions of top quark production for different values of W' boson masse at the Tevatron at 1.96TeV, transverse momentum distribution and pseudorapidity distributions.

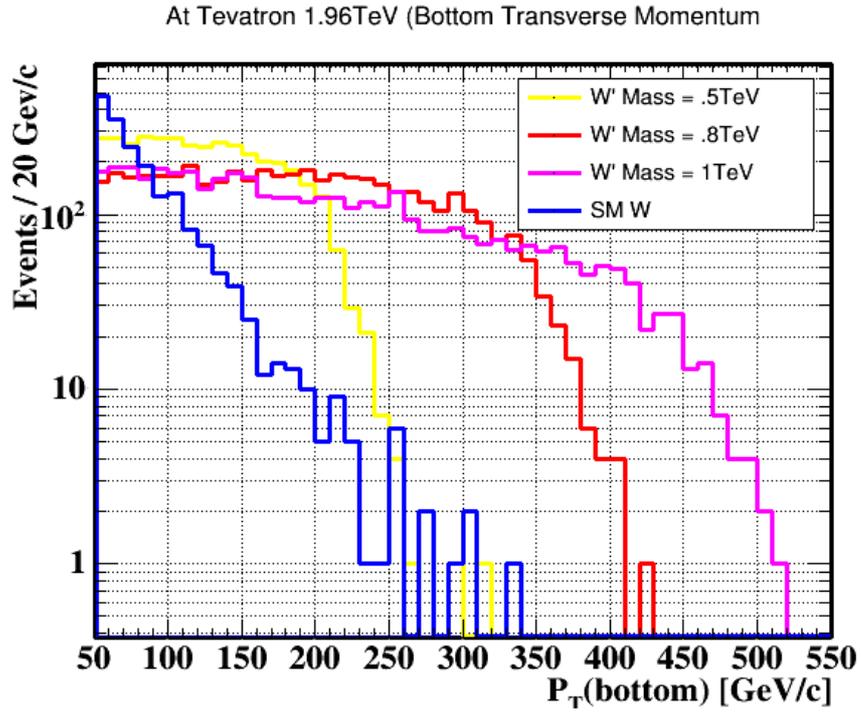

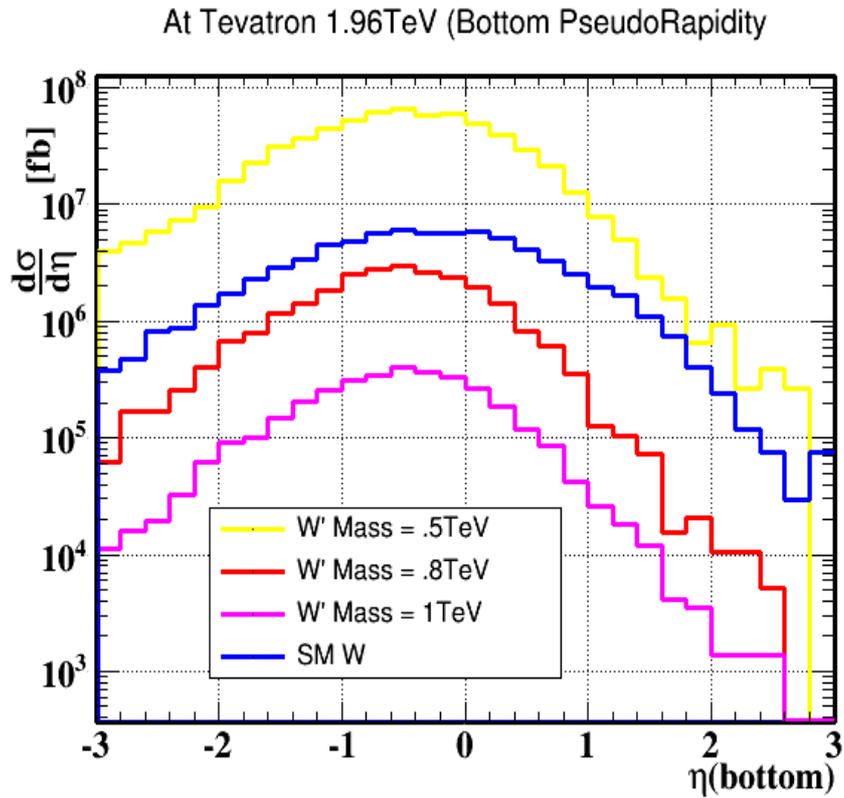

**Fig. 13:** The kinematics distributions of bottom quark production for different values of W' boson masse at the Tevatron at 1.96TeV, transverse momentum distribution and pseudorapidity distributions

# 1.6 Decay of top quark

Now, we will focus on the final state of single top quark production where the top quark decays into b quark and SM W boson, which subsequently decays leptonically, $W \to e\nu$ or $W \to \mu\nu$. This gives rise to an event signature with a high transverse momentum lepton and significant missing transverse energy from the neutrino, in association with two b-quark jets. Figure 14 show the leading order (LO) Feynman diagram for W′ boson production resulting in single top quark events. This diagram is identical to that for SM s-channel single top quark production

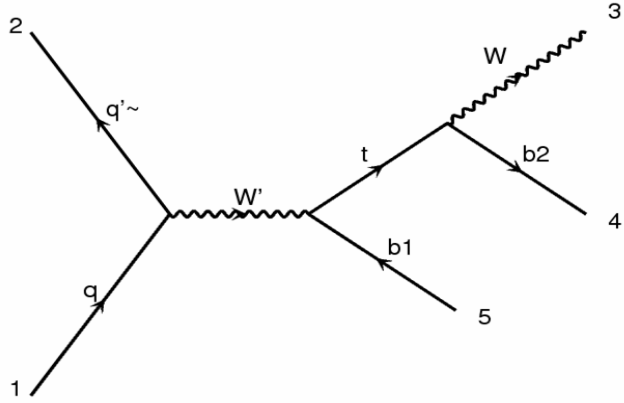

**FIG. 14:** Leading order Feynman diagram for single top quark production via heavy W′ boson, the top quark decays to a SM W boson and a b quark

We select signal-like events of top quark and separate the data into independent analysis sets based on final-state lepton flavor (electron or muon) and b-tag multiplicity (single tagged or double tagged), where b-quark jets are tagged using reconstructed displaced vertices in the jets. The independent datasets are later combined in the final statistical analysis. We performed an analysis on the invariant mass distribution of all final state objects and obtained the upper cross section limits at discrete W′ mass points and compared these limits to the theoretical prediction see figure [4]. The Monte Carlo data for this analysis are generated by MadGraph5 and MadEvents for the LHC from 6 TeV to 14 TeV proton-proton collider and for the Fermilab Tevatron, a 1.96 TeV proton-antiproton collider and showering and hadronization by Pythia8 see figures [5-8]

## III. Detection W' Signal via Charged Leptons

From the previous sections we searched for W' boson in its decay to a top quark and a bottom quark. In the Standard Model, the top quark decays to a W boson and b quark, where the W boson decays either hadronically, into a quark anti-quark pair, or leptonically, into a charged lepton and a neutrino. Now we investigate the leptonic decay channel of W boson of the Standard Model to detect the W' boson signal via the collision $pp(\bar{p}) \rightarrow W' \rightarrow t\bar{b} \rightarrow \ell\nu bb$ as shown in figure 15. The final state signature consists of two b quarks, one lepton (electron or muon) and missing transverse energy resulting from an undetected neutrino.

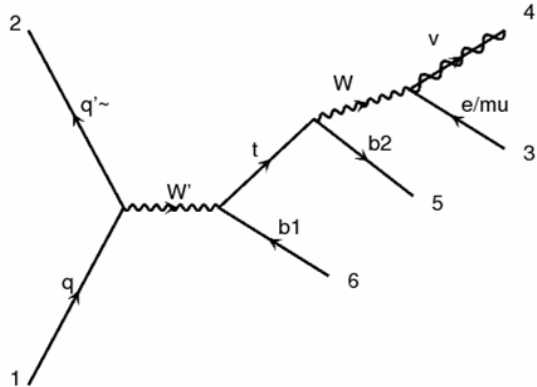

**Fig. 15:** The final state of W' boson decay where W boson decay leptonicaly to electron or muon plus neutrino

As we mentioned this analysis is based on the detection of events with a lepton (e, m), jets, and missing transverse energy in the final state. We analyzed MC events generated MadGraph5 contaib electrons or muons, jets, and missing transverse energy ($E_T^{miss}$). The $W' \rightarrow tb \rightarrow \ell\nu bb$ decay is characterized by the presence of a high $P_T$ isolated lepton, significant $E_T^{miss}$ associated with the neutrino, and at least two high $P_T$ b-jets. Monte Carlo (MC) techniques are used to model the W' signal in context of Left-Right Symmetric Model. We simulate W' bosons with mass values ranging from 0.5 to 5 TeV.

## 1. Kinematics Cuts at the LHC

We put kinematics cuts on samples events of electrons, muons, jets, and $E_T^{miss}$. Candidate events are required to pass an isolated electron with a pT >35 GeV and the electrons are required to have relative isolation less than $\Delta R < 0.125$ and for muon are required to have relative isolation $\Delta R < 0.15$ and pT>32 GeV. Event containing a second lepton with relative isolation $\Delta R < 0.2$ and a minimum pT requirement for muon 10 GeV and fo electron 15 GeV. Muon or electron is required to be separated from jets by $\Delta R(jet, l) > 0.3$ and the relative isolation for hadrons and photons $\Delta R < 0.4$. At least one lepton is required to be within the detector acceptance $|\eta| < 2.5$. For bottom quarks $\Delta R = 0.5$, pT >30 GeV and $|\eta| < 2.4$. The $E_T^{miss}$ is required to exceed in both the electron and muon samples in order to reduce the QCD multijet background. For missing energy with muon decay channel must be $E_T^{miss} > 20 GeV$ and with electron decay channel is $E_T^{miss} > 35 GeV$ .. Jets are required to be $P_T >$ 30 GeV and $|\eta| < 2.4$. At least two jets are required in the event with the highest-pT the first jet has pT > 120 GeV and the second jet pT > 40 GeV [26].

## 2. Kinematics cuts at the Tevatron

The selected events at Tevatron proton-antiproton collision in the electron channel, the events are selected by requiring one isolated electron with transverse momentum $p_T > 15 GeV$ and $|\eta| < 1.1$. In the muon channel, the events are selected by requiring one isolated muon with transverse momentum $p_T > 15 GeV$ and $|\eta| < 2$. For both channels, the events are also required to have missing transverse energy $E_T^{miss} = 15 GeV$ [27]. Jets are required to have $p_T > 15 GeV$ and $|\eta| < 3.4$. Events must have two jets, with the leading jet additionally required to have $p_T > 25 GeV$ and $|\eta| < 2.5$.

# 3. Production Cross Section at the LHC

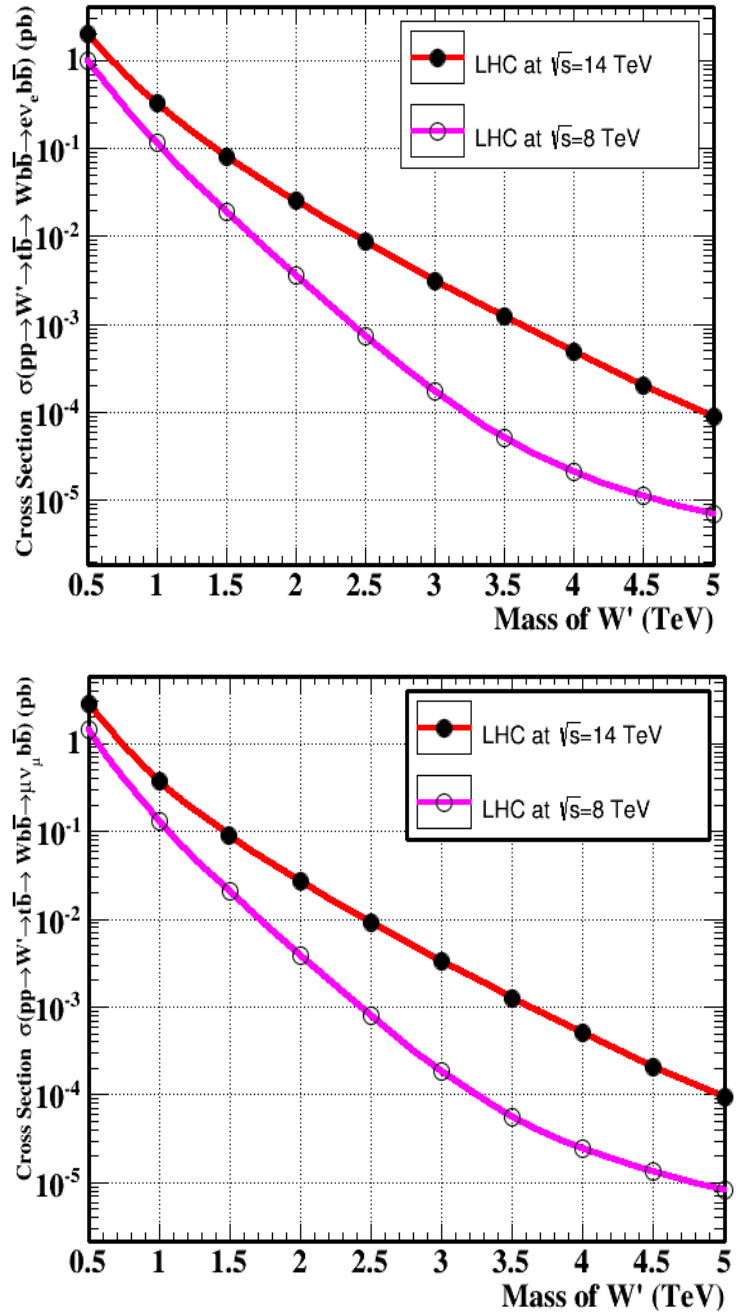

**Fig. 16:** The Production Cross Section of W' signal at the LHC when the center of mass energies are 8 TeV and 14 TeV via the decay channel of electron and missing energy final state (top), the decay channel of muon and missing energy final state (bottom) when W' mass is 1 TeV and top quark mass is 172 GeV

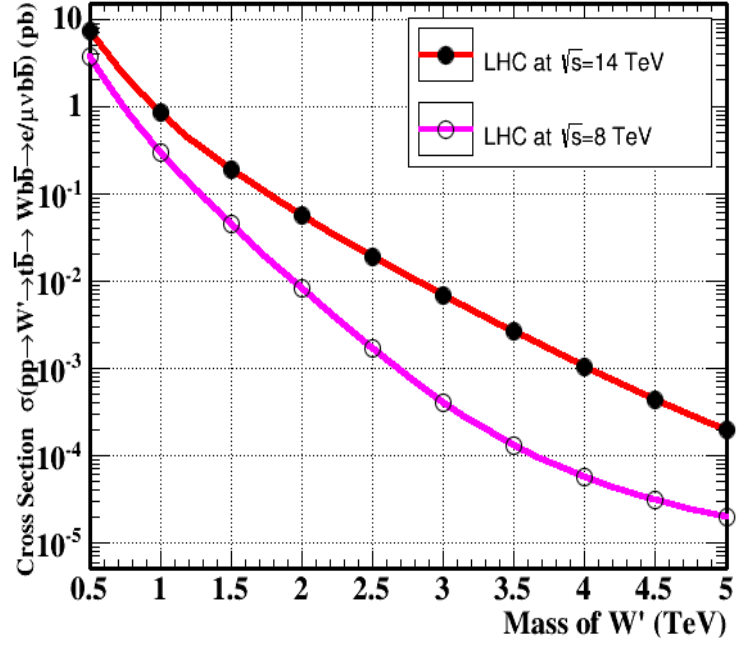

**Fig. 17:** The Production Cross Section of W' signal at the LHC when the center of mass energies are 8 TeV and 14 TeV via events sample of (electron and muon) and missing energy final state when W' mass is 1 TeV and top quark mass is 172 GeV

## 4. Production Cross Section at the Tevatron

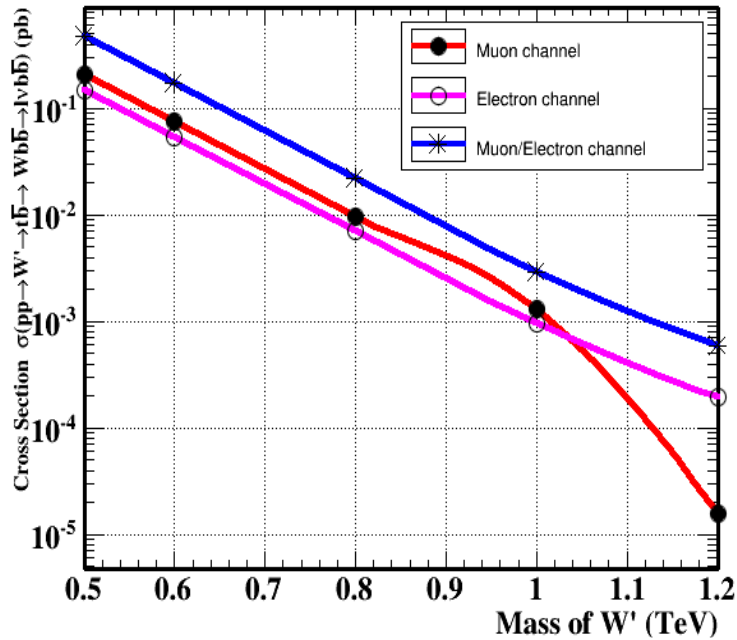

**Fig. 18:** The Production Cross Section of W' signal at the Tevatron when the center of mass energy is 1.96TeV via the decay channel of (electron, muon or both) and missing transverse energy in final state when W' mass is 1 TeV and top quark mass is 172 GeV

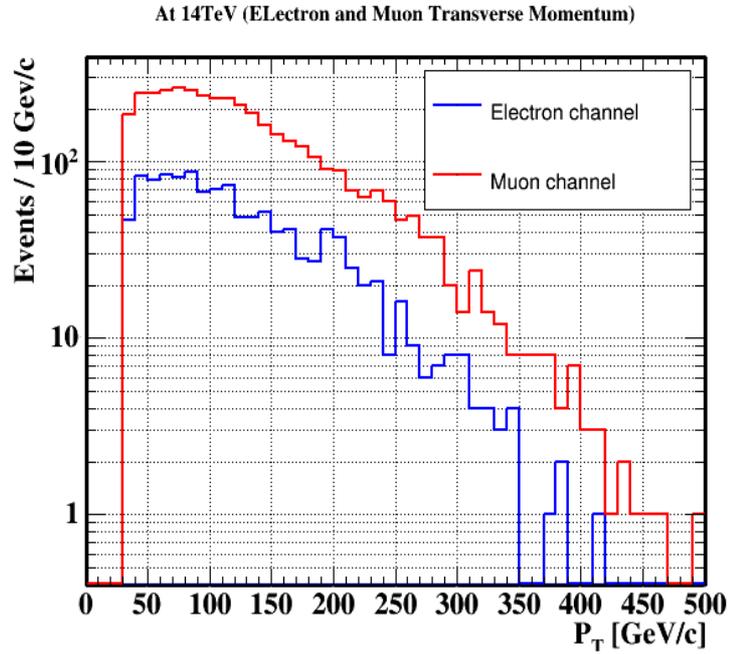

**Fig. 19:** Histogram shows the comparison between the transverse momentum of electron and muon of W' decay final state at the LHC when the center of mass energy is 14 TeV, the mass of W' is 1TeV and the mass of top is 172 GeV.

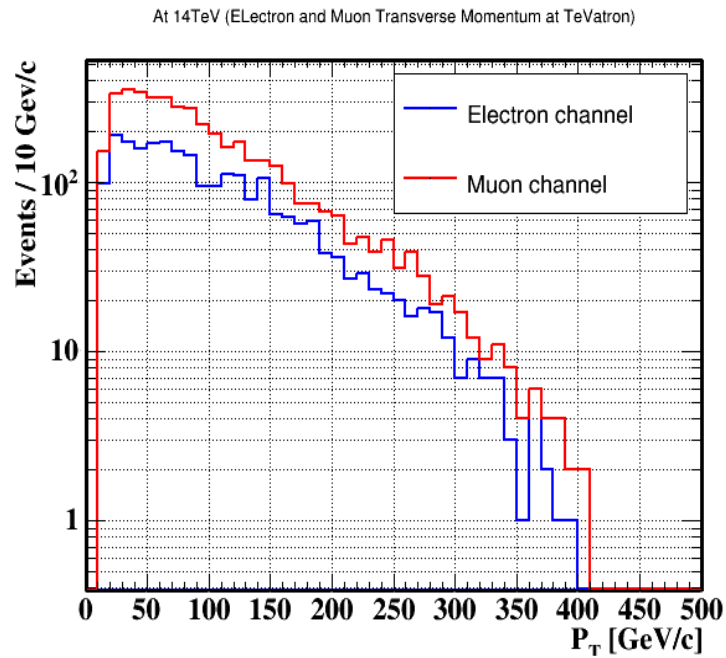

**Fig. 20:** Histogram shows the comparison between the transverse momentum of electron and muon of W' decay final state at the Tevatron when the center of mass energy is 1.96 TeV, the mass of W' is 1TeV and the mass of top is 172 GeV.

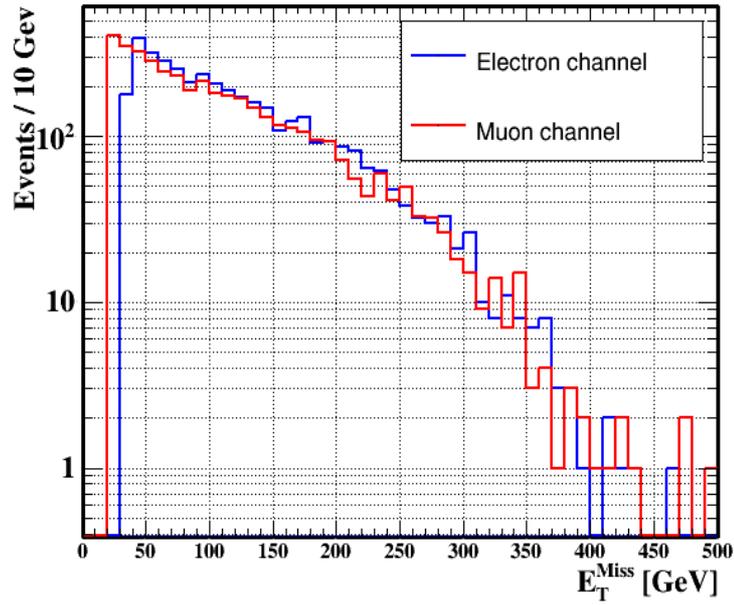

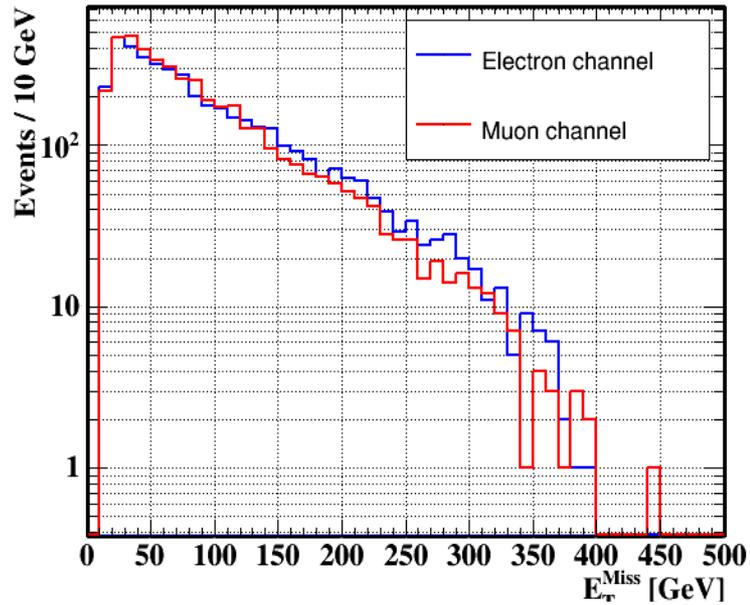

**Fig. 21:** Histograms show the comparison between the missing transverse energy in the final state produced from W' decay at the LHC when the center of mass energy is 14 TeV and at the Tevatron when the center of mass energy is 1.96 TeV , the mass of W' is 1TeV and the mass of top is 172 GeV.

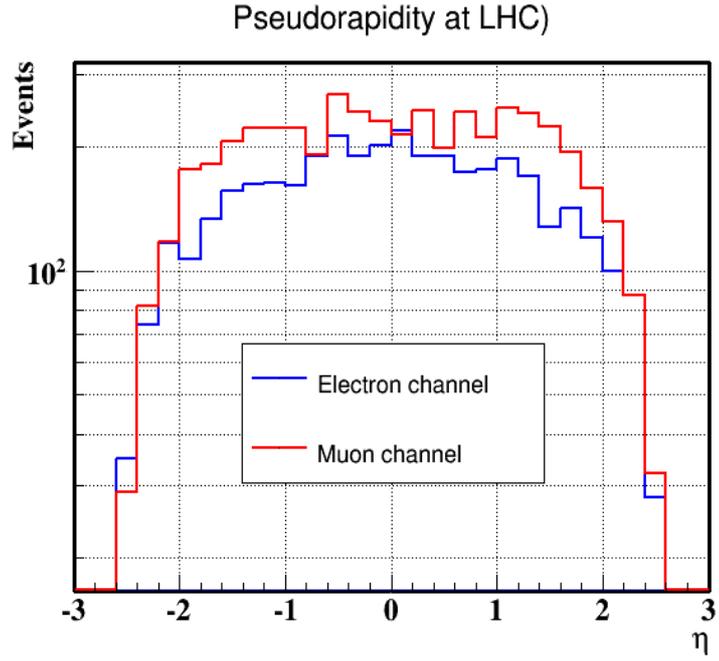

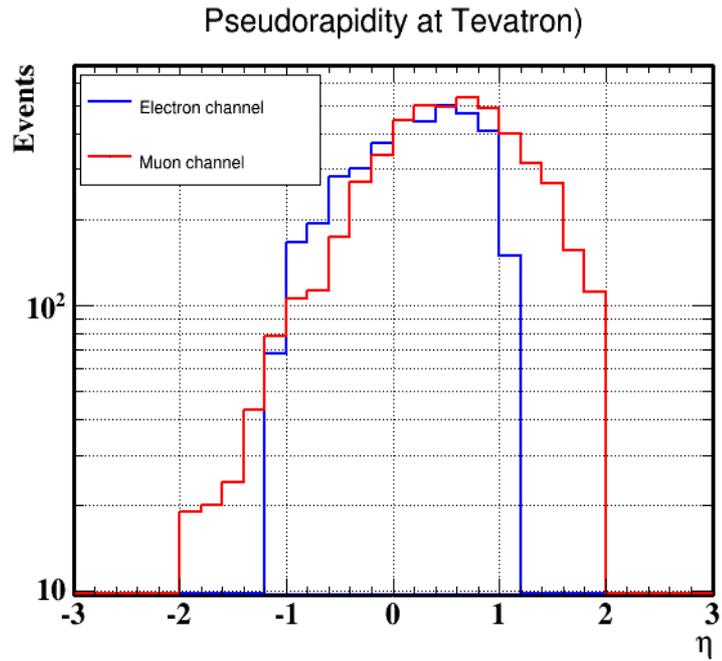

**Fig. 22:** Histograms shows the comparison between pseudorapidity of electron and muon in the final state produced from W' decay at the LHC when the center of mass energy is 14 TeV and at the Tevatron when the center of mass energy is 1.96 TeV , the mass of W' is 1TeV and the mass of top is 172 GeV.

# IV. Conclusion

In this work we presented an analysis for the potential discovery of the hypothetical W′ new charged heavy vector boson in single-top production process at the hadron colliders by analyzing the data generated by Monte Carlo event generators at the LHC and the Tevatron energies in the context of Left-Right Symmetric Model (LRSM) an extension of the Standard Model based on the gauge symmetry group SU(3)C × SU(2)L × SU(2)R × U(1)B-L .We calculated the production cross section for W' boson, different branching ratios of W' boson decay, the width of W' and reconstructed W' mass from the invariant masses of top and bottom quarks at different energies. Also we presented an analysis to detect the W' boson signal at hadron colliders via charged lepton (electron or muon) and missing transverse energy in final state $b\bar{b}l\nu$ produced from the collision $pp(\bar{p}) \rightarrow W' \rightarrow t\bar{b}$ ; $t \rightarrow Wb$ ; $W \rightarrow \ell\nu$ ; $\ell = e/\mu$ after applying the different kinematics cuts on top, bottom, charged lepton and missing transverse energy. We compared our measurements to the previous work theoretically and experimentally where from the previous researches the theoretical value of production cross section of W' boson is 15 pb and from our simulation and analysis for LRSM we calculated the production cross section of W' boson at the LHC at center of mass energy 10 TeV with mass 1TeV is 13.31 pb. Also this value is near from the current limit on the single top quark production cross section in the s-channel which is 13.6 pb . So we expect the production of top quark may be come also from the new charged heavy gauge boson W'.

# Acknowledgment

It is a pleasure to thank Prof. Torbjörn Sjostrand - Lund Univ. Sweden for useful discussions of PYTHIA, Prof. Johan Alwall - Stanford Univ. USA for useful discussions of MadGraph5/MadEvent. Also many thanks to Prof. Sherief Mourad, Prof. A. Abdelsalam , Prof. Omar Osman and Prof. Amr Radi,